\documentclass[12pt, a4paper]{article}
\usepackage{times}
\usepackage{amsmath}
\usepackage{amssymb}
\usepackage{amsthm}
\usepackage[dvips]{graphicx}
\usepackage{bm}
\usepackage{lscape}
\usepackage{leftidx}
\usepackage{ mathrsfs }
\usepackage[top=20mm,bottom=25mm,left=25mm,right=22mm]{geometry}
\usepackage{setspace}
\usepackage{cite}
\begin{document}

\vspace{5mm}
\hspace{-6mm}{\Large First-Principles Theory of the Momentum-Dependent Local Ansatz}

\vspace{3mm}
{
\hspace{35mm}{\Large  for Correlated Electron System}
\footnote{To be publiched in Physics Procedia}}

\vspace{5mm} 
\hspace{42mm} Sumal Chandra* and Yoshiro Kakehashi**

{
\hspace{32mm} University of the Ryukyus, Nishihara, Okinawa, Japan
}

{
\hspace{32mm}*k138609@eve.u-ryukyu.ac.jp,**yok@sci.u-ryukyu.ac.jp}

\vspace{10mm} 
\hspace{65mm} {\large Abstract}
\vspace{3mm}

\hspace{-6mm}The momentum-dependent local-ansatz (MLA) wavefunction describes well correlated electrons in solids in both the weak and strong interaction regimes. In order to apply the theory to the realistic system, we have extended the MLA to the first-principles version using the tight-binding LDA+U Hamiltonian. We demonstrate for the paramagnetic Fe that the first-principles MLA can describe a reasonable correlation energy gain and suppression of charge fluctuations due to electron correlations. Furthermore, we show that the MLA yields a distinct momentum dependence of the momentum distribution, and thus improves the Gutzwiller wavefunction.
\vspace{3mm}

\hspace{-6mm}{\it Keywords}: variational method, local-ansatz wavefunction, Gutzwiller wavefunction, electron correlations

\vspace{6mm} 
\hspace{-6mm}{\large 1\hspace{3mm}Introduction}
\vspace{3mm}

\hspace{-6mm}Electron correlations play an important role on the properties of materials such as the magnetism, the metal-insulator transition, and the high-temperature superconductivity. The wavefunction method is a simple and intuitive method to describe the correlated electrons leading to these properties. In this method, we construct a trial wavefunction with variational parameters choosing the minimum basis set in the Hilbert space to describe the correlated motion of electrons, and find the best wavefunction at the ground state on the basis of the variational principle.

The Gutzwiller wavefunction (GW)\cite{mcg63, mcg64, mcg65} is one of the oldest and useful trial wavefunctions in solids. The wavefunction describes the correlated electrons controlling the amplitudes of the doubly occupied states in the Hartree-Fock wavefunction. The GW has been applied to many phenomena such as the correlation effects on the magnetism, the heavy-fermion behavior, and the metal-insulator transition. The local-ansatz wavefunction (LA)\cite{gs77, gs78, gs80} is an alternative wavefunction. It takes into account the correlated states expanded in terms of the residual Coulomb interactions, which are neglected in the Hartree-Fock wavefunction. The LA has been applied to many systems such as the transition metals and the semiconductors\cite{pf95, pf12}.

The wavefunctions mentioned above, however, do not yield the exact results of the Rayleigh-Schr\"odinger perturbation theory in the weak Coulomb interaction limit. We therefore recently proposed the momentum-dependent local ansatz (MLA) wavefunction \cite{ykt08, mar11, mar13, yk14}. In the MLA, we first expand the Hilbert space from the Hartree-Fock ground state $\vert{\phi}_{0}\rangle$ by means of the two-particle excitation operators in the momentum representation. After introducing the momentum dependent amplitudes for the excitation operators as the variational parameters, we project these operators onto the local orbitals. Resulting correlators $\{\tilde{O}_{i}\}$ are more suitable than those created by the LA, for the description of electron correlations in solids. We also emphasize that the exact wavefunction in the weak interaction limit is essential to describe the quasiparticle weight in the strongly correlated regime, as shown in our paper \cite{mar11}, because the latter is obtained by the renormalization of that in the weak interaction limit in the variational approach. In the second paper \cite{mar13}, we extended the MLA to the strongly correlated regime introducing a hybrid wavefunction as a starting wavefunction. The potential of the hybrid wavefunction can change from the Hartree-Fock type to the alloy-analogy type by varying a weighting factor from zero to one, so that we can start from the best wavefunction for a given Coulomb interaction strength. We have shown on the basis of numerical calculations that the MLA improves the LA and the GW \cite{mar11, mar13}. 

In this paper, we propose the first-principles MLA combining the MLA with the first-principles tight-binding Hamiltonian, called the LDA (local-density approximation)+U Hamiltonian. The first-principles MLA allows us to describe the correlations of the multiband system, especially, the on-site charge-charge correlations, as well as the Hund-rule correlations at the ground states. We limit ourself in the present paper to the MLA in the metallic regime starting from the Hartree-Fock wavefunction. The description of the strongly correlated regime with use of a hybrid wavefunction will be presented elsewhere. 

In the following section, we propose the first-principles MLA wavefunction, and obtain the correlation energy in the single-site approximation (SSA). Using the variational principle, we obtain an approximate form of the variational parameters.\! In Sec.\,3, we present some numerical results of calculations for paramagnetic Fe in the weak Coulomb interaction regime. We verify the correlation energy gain, the suppression of charge fluctuations, and in particular the momentum dependent behavior of the momentum distribution. In the last section 4, we summarize the present work.

\vspace{6mm} 
\hspace{-6mm}{\large 2 \hspace{3mm}LDA+U Hamiltonian and MLA Wavefunction}
\vspace{3mm}

\hspace{-6mm}We consider here the transition metal  system for simplicity and adopt the first-principles LDA+U Hamiltonian, which is based on the tight-binding linear muffin-tin orbital method \cite{via10, yk12}.
\begin{equation}
H=H_{0}+H_{1},
\label{equ1}
\end{equation}
\begin{equation}
H_{0}=\sum_{iL\sigma}\epsilon^{0}_{iL} \hat{n}_{iL\sigma} +\sum_{iLjL^{'}\sigma}{t}_{iLjL^{'}} a^{\dagger}_{iL\sigma} {a}_{jL^{'}\sigma},
\label{equ2}
\end{equation}
\begin{align}
\!H_{1}= \!\sum_{i}\left[\sum_{m} {U}_{mm} \,\hat{n}_{ilm\uparrow}\hat{n}_{ilm\downarrow}+\!\!\sum_{m>m'} \!\! \left(U_{mm'}-\frac{1}{2}J_{mm'}\!\!\right)\hat{n}_{ilm} \hat{n}_{ilm{'}}-\!\!\!\!\sum_{m>m'}\!\! J_{mm'} \,\hat{\bm {s}}_{{ilm}}\cdot\hat{\bm {s}}_{ilm'}\right].
\label{equ3}
\end{align}
Here $H_{0}$ and $H_{1}$ denote the non-interacting and interacting parts of the Hamiltonian $H$, respectively. $\epsilon^{0}_{iL}$ is the atomic level of the orbital $ L$ on site $i$. ${t}_{iLjL'}$  is the transfer integral between $iL$ and $jL'$. $L=(l, m)$ denotes the $s, p, d $ orbitals. $a^{\dagger}_{iL\sigma}{({a}_{iL\sigma})}$ is the creation (annihilation) operator for an electron on site $i$ with orbital $L$ and spin ${\sigma }$, and ${\hat{ n}_{iL\sigma}}=a^{\dagger}_{iL\sigma}{a}_{iL\sigma}$ ($\hat{\bm {s}}_{{iL}}=\sum_{\alpha \gamma} a^{\dagger}_{iL\alpha} (\bm{\sigma})_{\alpha\gamma}\,{a}_{iL\gamma}/2$) is the charge (spin) density operator. Here $\bm{\sigma}$ denote the Pauli spin matrices.
In the LDA+U Hamiltonian we assume that the $s p$ electrons are well described by the LDA in the band theory, and take into account only on-site Coulomb interactions between $d\, (l = 2)$ electrons in the interacting part $H_{1}$. $U_{mm} (U_{mm'} )$ and $J_{mm'}$ denote the intra-orbital (inter-orbital) Coulomb and exchange interactions, respectively. $\hat{n}_{ilm} (\hat{\bm {s}}_{ilm})$ with $l = 2 $ is the charge (spin) density operator for $d $ electrons on site $i$ and orbital $m$. The atomic level $\epsilon^{0}_{iL}$ in $H_{0}$ is calculated from the LDA atomic level $\epsilon_{iL}$ by subtracting the double counting potential as $\epsilon^{0}_{iL} = \epsilon_{iL} - \partial{E^{\mathrm{U}}_{\mathrm{ LDA}}}/\partial{n_{iL\sigma}}$. Here $n_{iL\sigma}$ is the charge density at the ground state, $E^{\mathrm{U}}_{\mathrm{ LDA}}$ is a LDA functional for the  intra-atomic Coulomb interactions. 

In the first-principles MLA, we rewrite the Hamiltonian $H$ as the sum of the Hartree-Fock Hamiltonian $H_{\mathrm{HF}}$ and the residual interactions $H_{\mathrm {I}}$. The latter is given by
\begin{equation}
H_{\mathrm{I}}=\sum_{i}{\left[\sum_{m}U_{mm}{O}^{(0)}_{imm}+\sum_{m>m'}\left(U_{mm'}-\frac{1}{2}J_{mm'}\right){O}^{(1)}_{imm'}-\sum_{m>m'}J_{mm'}{O}^{(2)}_{imm'}\right]}.
\label{equ4}
\end{equation}
Here ${O}^{(0)}_{imm}$, ${O}^{(1)}_{imm'}$, and ${O}^{(2)}_{imm'}$ denote the two-particle  operators: the intra-orbital operators, the charge-charge inter-orbital operators, and the spin-spin inter-orbital operators, respectively, and are defined as follows.

\begin{equation}
{O}^{(0)}_{imm}=\delta\hat{n}_{ilm\uparrow}\delta\hat{n}_{ilm\downarrow},
\label{equ5}
\end{equation}

\begin{equation}
{O}^{(1)}_{imm'}=\delta\hat{n}_{ilm}\delta\hat{n}_{ilm'},
\label{equ6}
\end{equation}

\begin{equation}
{O}^{(2)}_{imm'}=\delta\hat{\bm{s}}_{ilm}\cdot\delta\hat{\bm{s}}_{ilm'}.
\label{equ7}
\end{equation}
Here $\delta n_{ilm\sigma}=n_{ilm\sigma}-\langle n_{ilm\sigma}\rangle_{0}$, $\delta n_{ilm}=n_{ilm}-\langle n_{ilm}\rangle_{0}$, and $\delta \bm{s}_{ilm}=\bm{s}_{ilm}-\langle \bm{s}_{ilm}\rangle_{0}$,\, $\langle \sim \rangle_{0}$ denotes the average in the Hartree-Fock approximation.

In the residual interactions (\ref{equ4}), we have three types of correlation operators, $O^{(0)}_{imm}$, ${O}^{(1)}_{imm'}$, and ${O}^{(2)}_{imm'}$. Accordingly, we construct the MLA wavefunction as follows.
\begin{equation}
\vert{\Psi}_\mathrm{MLA}\rangle={\left[\prod_{i}{\left( 1-\sum_{m}{\tilde{O}}^{(0)}_{imm}-\sum_{m>m'}{\tilde{O}}^{(1)}_{imm'}-\sum_{m>m'}{\tilde{O}}^{(2)}_{imm'}\right) } \right]}\vert{\phi}_{0}\rangle.
\label{equ13}
\end{equation}
%
The correlators $\tilde{O}^{(0)}_{imm}$, $\tilde{O}^{(1)}_{imm'}$, and $\tilde{O}^{(2)}_{imm'}$ are the two-particle operators projected onto the local orbitals, and describe the intra-orbital correlations, the charge-charge inter-orbital correlations, and the spin-spin inter-orbital correlations, respectively. They are defined by
\begin{align}
\tilde{O}^{(q)}_{iLL'}&= \sum_{\{kn\sigma\}}\langle{k'_{2}n'_{2}\vert iL}\rangle \langle{iL\vert {k}_{2}{n}_{2}}\rangle \langle{k'_{1}n'_{1}\vert iL'}\rangle \langle{iL'\vert {k}_{1}{n}_{1}}\rangle  \nonumber \\
              &\quad\qquad\times\lambda^{(q)}_{{LL'}\{{2'2 1'1}\}}\delta(a^{\dagger}_{k'_{2}n'_{2}\sigma'_{2}}a_{{k}_{2}{n}_{2}\sigma_{2}})\delta(a^{\dagger}_{k'_{1}n'_{1}\sigma'_{1}}a_{{k}_{1}{n}_{1}\sigma_{1}}).
\label{equ9}
\end{align}
Here $q=$ 0, 1, and 2. $a^{\dagger}_{k n\sigma}{(a_{kn\sigma})}$ is the creation (annihilation) operator for an electron with momentum $k$, band index $n$, and spin $\sigma$. They are given by those in the site representation as $a_{kn\sigma}=\sum_{iL}a_{iL\sigma}\langle kn\vert iL\rangle$. 

The momentum dependent amplitudes $\lambda^{(q)}_{{LL'}\{{2'2 1'1}\}}$ in Eq. (\ref{equ9}) are given by

\begin{equation}
\qquad\qquad\lambda^{(0)}_{{LL'}\{{2'2 1'1}\}}=\eta_{L k'_{2}n'_{2}k_{2}n_{2}k'_{1}n'_{1}k_{1}n_{1}}\delta_{LL'}\delta_{\sigma'_{2}\downarrow}\delta_{\sigma_{2}\uparrow}\delta_{\sigma'_{1}\downarrow}\delta_{\sigma_{1}\uparrow},\nonumber \\
\label{equ10}
\end{equation}

\begin{equation}
\lambda^{(1)}_{{LL'}\{2'2 1'1\}}=\zeta^{(\sigma_{2}\sigma_{1})}_{L L'k'_{2}n'_{2}k_{2}n_{2}k'_{1}n'_{1}k_{1}n_{1}}\delta_{\sigma'_{2}\sigma_{2}} \delta_{\sigma'_{1}\sigma_{1}},\nonumber \\
\label{equ11}
\end{equation}

\begin{align}
\quad\qquad\qquad\lambda^{(2)}_{{LL'}\{{2'2 1'1}\}}&=\sum_{\sigma}\xi^{(\sigma)}_{L L'k'_{2}n'_{2}k_{2}n_{2}k'_{1}n'_{1}k_{1}n_{1}}\delta_{\sigma'_{2}-\sigma} \delta_{\sigma_{2}\sigma}\delta_{\sigma'_{1}-\sigma} \delta_{\sigma_{1}\sigma} \nonumber \\
                                  &\quad+\frac{1}{2} \sigma_{1} \sigma_{2} \,  \xi^{(\sigma_{2}\sigma_{1})}_{L L'k'_{2}n'_{2}k_{2}n_{2}k'_{1}n'_{1}k_{1}n_{1}}\delta_{\sigma'_{2}\sigma_{2}} \delta_{\sigma'_{1}\sigma_{1}}.                    
\label{equ12}
\end{align}
 Here $\eta_{L k'_{2}n'_{2}k_{2}n_{2}k'_{1}n'_{1}k_{1}n_{1}}$, $\zeta^{(\sigma_{2}\sigma_{1})}_{L L'k'_{2}n'_{2}k_{2}n_{2}k'_{1}n'_{1}k_{1}n_{1}}$, $\xi^{(\sigma)}_{L L'k'_{2}n'_{2}k_{2}n_{2}k'_{1}n'_{1}k_{1}n_{1}}$, and $\xi^{(\sigma_{2}\sigma_{1})}_{L L'k'_{2}n'_{2}k_{2}n_{2}k'_{1}n'_{1}k_{1}n_{1}}$ are the variational parameters. 

It should be noted that the MLA wavefunction (\ref{equ13}) describes exactly the weak interaction regime. The best wavefunction is obtained by choosing the variational parameters best on the basis of the variational principles for the ground state energy.

The correlation energy is defined by $\langle H \rangle-\langle H \rangle_{0}$, where $\langle \sim \rangle$ ($\langle \sim \rangle_{0}$) denotes the full (Hartree-Fock) average. The MLA contains an infinite number of variational parameters, so that the advanced numerical methods such as the variational Monte-Carlo technique are not applicable for the calculation of the energy. We adopt here the single-site approximation (SSA). The correlation energy per atom $\epsilon_c$ is then obtained as follows.
 \begin{equation}
{\epsilon_c} =\frac{{- \langle {\tilde{O_i}^\dagger}} \tilde{H} \rangle_0 
- \langle \tilde{H} \tilde{O_i}\rangle_0 
+\langle {\tilde{O_i}^\dagger} \tilde{H} \tilde{O_i}\rangle_0 }{1+\langle{\tilde{O_i}^\dagger\tilde{O_i}}\rangle_0}. 
\label{equ14}
\end{equation}
Here $ \tilde{H}=H-\langle H\rangle_0 $, and 
\begin{equation}
 \tilde{O_i}=\sum_{m}{\tilde{O}}^{(0)}_{imm}+\sum_{m>m'}{\tilde{O}}^{(1)}_{imm'}+\sum_{m>m'}{\tilde{O}}^{(2)}_{imm'}.
\label{equa15}
\end{equation}

Solving the self-consistent equations obtained from the stationary condition $\delta\epsilon_{c}=0$, we find an approximate form of variational parameters in the weak Coulomb interaction regime as follows.
\begin{equation}
\eta_{L k'_{2}n'_{2}k_{2}n_{2}k'_{1}n'_{1}k_{1}n_{1}}=\frac{U_{mm}\tilde{\eta}_{m}}{\Delta E_{{k'_{2}n'_{2}\downarrow} k_{2}n_{2}\downarrow k'_{1}n'_{1}\uparrow_{1} k_{1}n_{1}\uparrow}-\epsilon_c},
\label{equ16}
\end{equation}
\begin{equation}
\zeta^{(\sigma\sigma')}_{L L'k'_{2}n'_{2}k_{2}n_{2}k'_{1}n'_{1}k_{1}n_{1}}=\frac{(U_{mm'}-J_{mm'}/2)\tilde\zeta^{(\sigma\sigma')}_{mm'}}{\Delta E_{{k'_{2}n'_{2}\sigma'} k_{2}n_{2}\sigma k'_{1}n'_{1}\sigma' k_{1}n_{1}\sigma}-\epsilon_c},
\label{equ17}
\end{equation}
\begin{equation}
\quad\xi^{(\sigma)}_{L L' k'_{2}n'_{2}k_{2}n_{2}k'_{1}n'_{1}k_{1}n_{1}}=\frac{J_{mm'}\tilde \xi^{(\sigma)}_{mm'}}{\Delta E_{{k'_{2}n'_{2}{-\sigma}} k_{2}n_{2}\sigma k'_{1}n'_{1}\sigma k_{1}n_{1}{-\sigma}}-\epsilon_c},
\label{equ18}
\end{equation}
\begin{equation}
\xi^{(\sigma\sigma')}_{L L'k'_{2}n'_{2}k_{2}n_{2}k'_{1}n'_{1}k_{1}n_{1}}=\frac{J_{mm'}\tilde \xi^{(\sigma\sigma')}_{mm'}}{\Delta E_{{k'_{2}n'_{2}\sigma'} k_{2}n_{2}\sigma k'_{1}n'_{1}\sigma' k_{1}n_{1}\sigma}-\epsilon_c}.
\label{equ19}
\end{equation}
 Here $\Delta E_{k'_{2}n'_{2}\sigma'_{2} k_{2}n_{2}\sigma_{2} k'_{1}n'_{1}\sigma'_{1} k_{1}n_{1}\sigma_{1}}= \epsilon_{k'_{2}n'_{2}\sigma'_{2}}-\epsilon_{k_{2}{n}_{2}\sigma_{2}}+\epsilon_{k'_{1}n'_{1}\sigma{'}_{1}}-\epsilon_{k_{1}{n}_{1}\sigma_{1}}$ denotes  the two-particle excitation energy. $\tilde{\eta}_{m}$, $\tilde\zeta^{(\sigma\sigma')}_{mm'}$, $\tilde \xi^{(\sigma)}_{mm'}$, and $\tilde \xi^{(\sigma\sigma')}_{mm'}$ are the variational parameters. Substituting Eqs.(\ref{equ16})$\sim $(\ref{equ19}) into Eq.(\ref{equ14}) and using the variational principles $\delta\epsilon_{c}=0$ again, we obtain the self-consistent equations for the variational parameters to be solved. The elements $\langle {\tilde{O_i}^\dagger} \tilde{H}\rangle_0 $, $\langle {\tilde{O_i}^\dagger} \tilde{H} \tilde{O_i}\rangle_0 $, and $\langle {\tilde{O_i}^\dagger} \tilde{O_i}\rangle_0 $ in Eq.(\ref{equ14}) are calculated by Wick's theorem.

The other physical quantities such as the electron number and charge fluctuations are also obtained by taking average with respect to the MLA wavefunction (\ref{equ13}) and making the SSA.
%


\vspace{6mm} 
\hspace{-6mm}{\large 3\hspace{3mm}Numerical Results in the  Weak Interaction Regime}
\vspace{3mm}

\hspace{-6mm}The ferromagnetism and related properties of Fe have extensively been investigated theoretically with use of realistic Hamiltonians with $s$, $p$, and $d$ orbitals at the ground states\cite{xy09, omi08} and at finite temperatures \cite{ail01, yka11, mni13}. But its physics has not yet been fully clarified. We performed numerical calculations for the paramagnetic Fe in order to clarify the basic behavior of the first-principles MLA in the weak Coulomb interaction regime. In the weak Coulomb and exchange energy limit, we obtain $\tilde{\eta}_{m}\rightarrow 1$, $\tilde\zeta^{(\sigma\sigma')}_{mm'}\rightarrow 1$, $\tilde \xi^{(\sigma)}_{mm'}\rightarrow -1$, and $\tilde \xi^{(\sigma\sigma')}_{mm'}\rightarrow -1$ from the self-consistent equations for the variational parameters. We adopted in the present calculations these variational parameters, and calculated physical quantities of Fe. Furthermore we neglected the orbital dependence in the Coulomb and exchange integrals in the numerical calculations, and adopted the values $U_{mm}=U_{0}=0.2749$ Ry, $U_{mm'}=U_{1}=0.1426$ Ry, and $J_{mm'}=J=0.0662$ Ry which are obtained from the relations $U_{0}=\bar{U}+8 \bar{J}/5, U_{1}=\bar{U}-2\bar{J}/5,$ and $J=\bar{J}$, using the average values $\bar{U}=0.1691$ Ry and $\bar{J}=0.0662$ Ry by Anisimov {\it{et al.}}\cite{via97}. Note that we adopted here the relation $U_{0}=U_{1}+2J$ for cubic system. The transfer integrals and the atomic level have been calculated with use of the Stuttgart tight-binding LMTO (linear muffin-tin orbital) package.
\begin{figure}[htbp]
\begin{center}
\includegraphics[scale=0.44, angle=270]{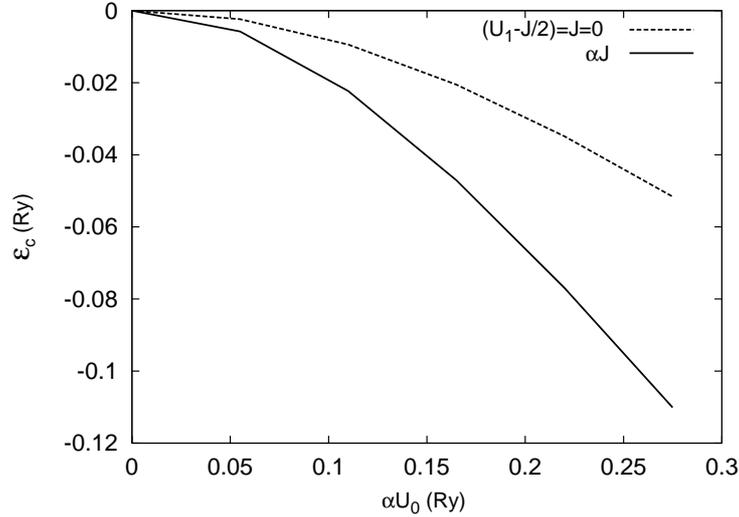}%
\caption{\label{fig1}
The correlation energy $\epsilon_c$ as a function of Coulomb interaction strength $\alpha U_{0}$ for the paramagnetic Fe. Dashed curve: the result without inter-orbital correlations ($i.e., U_{1}-J/2=J=0$), solid curve: the result with both the intra- and inter-orbital correlations.}
\end{center}
\end{figure}
\begin{figure}[htbp]
\begin{center}
\includegraphics[scale=0.44, angle=270]{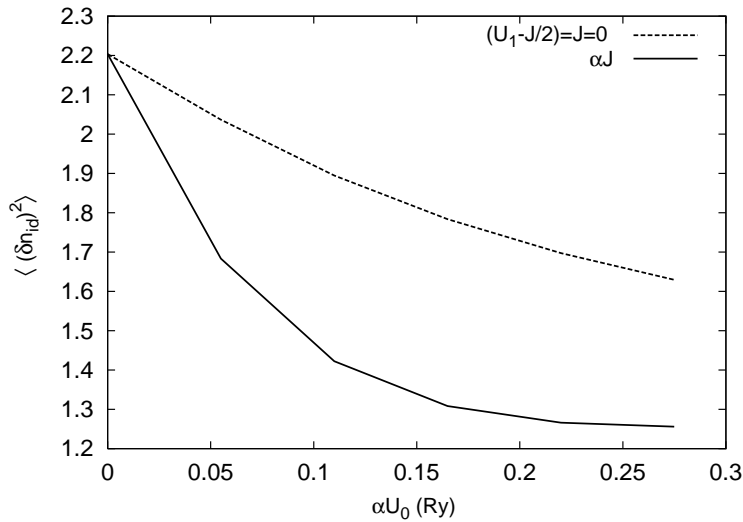}%
\caption{\label{fig2}
The charge fluctuation $\langle (\delta n_{id})^2\rangle$ vs Coulomb interaction strength $\alpha U_{0}$ curve for the paramagnetic Fe. Dashed curve: the result without inter-orbital correlations, solid curve: the result with both the intra- and inter-orbital correlations.}
\end{center}
\end{figure}

In order to see a systematic change due to correlation strength, we scaled $U_{0}$, $U_{1}$, and $J$ as $\alpha U_{0}$, $\alpha U_{1}$, and $\alpha J$ using a scaling factor $\alpha$.
Figure \ref{fig1} shows the calculated correlation energy as a function of $\alpha U_{0}$. With increasing $\alpha U_{0}$ (as well as $\alpha U_{1}$ and $\alpha J$), we find that the correlation energy $\epsilon_c$ monotonically decrease  and find $\epsilon_c = -0.0516$ Ry when $\alpha =1$ and $U_{1}=J=0$. When we take into account the inter-site correlations, the correlation energy $\epsilon_c$ decreases further and we obtain $\epsilon_c = -0.1101$ Ry when $\alpha =1$.

The correlation energy gain is accompanied by the suppression of charge fluctuations. We calculated the charge fluctuations for $d$ electrons $\langle (\delta n_{id})^2\rangle=\langle  n_{id}^2\rangle- \langle n_{id}\rangle^2$ as a function of $\alpha U_{0}$. As shown in Fig.\,\ref{fig2}, the charge fluctuation in the Hartree-Fock approximation is 2.2. The intra-orbital correlations suppress the charge fluctuations and yields $\langle (\delta n_{id})^2\rangle$=1.73 for $\alpha U_{0}=0.2$ Ry. The inter-orbital correlations more rapidly decrease the charge fluctuation with increasing $\alpha U_{0}$ as seen in Fig.\,\ref{fig2}. Calculated charge fluctuation is $\langle (\delta n_{id})^2\rangle \approx 1.3$ for $\alpha U_{0}=0.2$ Ry. The result is comparable to the value of the LA with the $d$-band model\cite{pf95}, $\langle (\delta n_{id})^2\rangle \approx 1.0$, but is somewhat larger than that of the LA because the present theory takes into account the hybridization between the $d$ and $sp$ electrons.
\begin{figure}[htbp]
\begin{center}
\includegraphics[scale=0.45, angle=270]{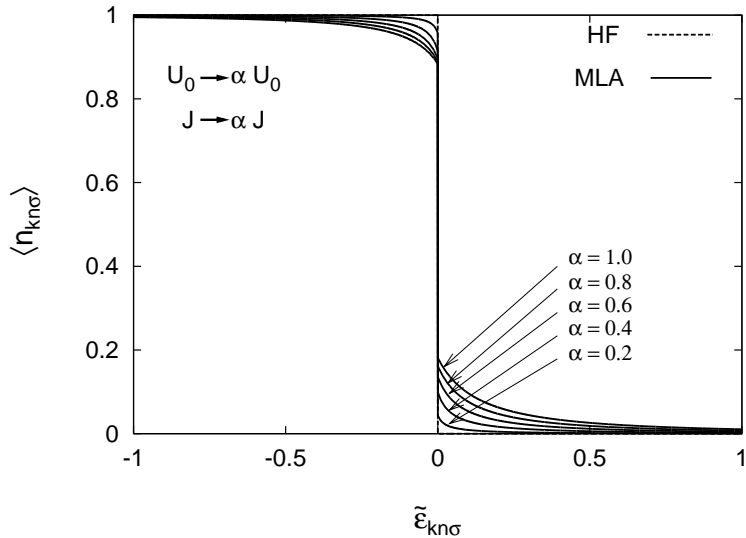}%
\caption{\label{fig3}
The momentum distribution $\langle n_{kn\sigma}\rangle$ as a function of the energy $\tilde{\epsilon}_{kn\sigma}$ for various scaling factors $\alpha$ of the Coulomb and exchange energies. Dash line denotes the distribution in the Hartree-Fock approximation.}
\end{center}
\end{figure}

We emphasized in the previous paper for the single-band model\cite{yk14} that the MLA can describe the momentum dependence of the momentum distribution, while those calculated from the original LA and the GW remain almost constant below and above the Fermi level. This behavior is seen also in the first-principles case. In fact, the momentum distribution $\langle n_{kn\sigma}\rangle$ is given in the present theory as follows. 
\begin{equation}
\langle n_{kn\sigma}\rangle=\langle n_{kn\sigma}\rangle_{0}+\frac{N\langle{\tilde{O_i}^\dagger\tilde{n}_{kn\sigma}\tilde{O_i}\rangle_{0}}}{1+\langle{\tilde{O_i}^\dagger\tilde{O_i}}\rangle_0}.
\label{equ20}
\end{equation}
The first term at the rhs (right-hand-side) is the momentum distribution in the Hartree-Fock approximation, which is given by the Fermi distribution function at zero temperature $f(\tilde{\epsilon}_{kn\sigma})=\theta (-\tilde{\epsilon}_{kn\sigma})$. Here $\theta$ denotes the step function, and $\tilde{\epsilon}_{kn\sigma}$ is the Hartree-Fock one-electron energy measured from the Fermi level. The second term at the rhs of  Eq. (\ref{equ20}) describes the correlation correction. $N$ denotes the number of atoms, $\tilde{n}_{kn\sigma}$ is defined by $\tilde{n}_{kn\sigma}= n_{kn\sigma}-\langle n_{kn\sigma}\rangle_{0}$.  The correlation correction consists of the terms being proportional to $\vert u_{Ln\sigma}(k)\vert^2$ $f(\tilde{\epsilon}_{kn\sigma})$  and those being proportional to $\vert u_{Ln\sigma}(k)\vert^2$ $f(-\tilde{\epsilon}_{kn\sigma})$. Here $\{ u_{Ln\sigma}(k)\}$  are the eigenvectors for given $k$ points. If we replace $\vert u_{Ln\sigma}(k)\vert^2$ with 1/5 as a rough approximation for the $d$-like branch  $n$ near the Fermi level, we find that $\langle n_{kn\sigma}\rangle$ depends on the momentum only via the energy $\tilde{\epsilon}_{kn\sigma}$ as in the single-band model. Figure \ref{fig3} shows the calculated result of momentum distributions in this approximation. We find clearly the momentum dependence of $\langle n_{kn\sigma}\rangle$ via $\tilde{\epsilon}_{kn\sigma}$, and obtain the mass enhancement $m^{*}=1.4$ for $\alpha=1.0$ from a jump at the Fermi level. This value should be compared with the experimental renormalization values 1.38 $\sim$ 2.12 which are obtained from the comparison of the LDA band calculations with the $T$-linear specific heat at low temperatures \cite{wp01, xy09}. We have to calculate $\langle n_{kn\sigma}\rangle$ along the high-symmetry lines taking into account the $k$-dependence of $\vert u_{Ln\sigma}(k)\vert^2$ in more detailed calculations. The results of the full calculations for $\langle n_{kn\sigma}\rangle$ will be published elsewhere.
%

\vspace{6mm} 
\hspace{-6mm}{\large 4\hspace{3mm}Summary}
\vspace{3mm}

\hspace{-6mm}We have extended the MLA to the first-principles MLA on the basis of the tight-binding LDA +U Hamiltonian. The wavefunction is constructed by applying the momentum-dependent intra-orbital correlators, the inter-orbital charge-charge correlators, and the spin-spin correlators to the Hartree-Fock uncorrelated wavefunction. The theory yields the exact results in the weak Coulomb interaction limit, and describes the charge-charge correlations as well as the Hund rule correlations in the real system.

By means of the numerical calculations for the paramagnetic Fe in the weak Coulomb interaction regime, we verified the correlation energy gain due to the inter-orbital charge-charge correlations as well as the spin-spin correlations between the $d$ orbitals. The charge fluctuation $\langle (\delta n_{id})^2\rangle$ in the present calculations is 1.3. The value is somewhat larger than the value obtained by the LA with the $d$-band model because of the hybridization between $sp$ and $d$ orbitals. We also clarified that the present theory leads to a distinct momentum dependence of the momentum distribution. The result qualitatively differs from those obtained from the LA and GW. The calculated effective mass $m^{*}=1.4$ is consistent with the experimental values obtained from the low-temperature specific heat data. Numerical calculations in the present paper are limited to the weak Coulomb interaction regime. In order to examine the quantitative aspects of the properties of correlated electrons in Fe and its compounds, we have to solve the full self-consistent equations for the variational parameters ($i.e.,$ $\tilde{\eta}_{m}$, $\tilde\zeta^{(\sigma\sigma')}_{mm'}$, $\tilde \xi^{(\sigma)}_{mm'}$, and $\tilde \xi^{(\sigma\sigma')}_{mm'}$). Numerical calculations for the correlated electron regime are in progress. 

\vspace{6mm} 
\hspace{-6mm}{\large  \hspace{3mm} Acknowledgement}
\vspace{3mm}

\hspace{-6mm}The present work is supported by a Grant-in-Aid for Scientific Research (25400404). The calculations have been partly performed by using the facilities of the Supercomputer Center, the Institute for Solid State Physics, University of Tokyo.

\end{document}